# On-chip coherent beam combination of waveguide amplifiers on $Er^{3+}$-doped thin film lithium niobate

*Rui Bao,[1,2] Lvbin Song,[1,2] Zhiwei Fang,[1,]\* Jinmin Chen,[1] Zhe Wang,[1] Jian Liu,[1] Lang Gao,[3] Zhaoxiang Liu,[1] Zhihao Zhang,[1] Min Wang,[1] Haisu Zhang,[1] and Ya Cheng[1,2,3,4,5,6,7,]\**

Corresponding author: zwfang@phy.ecnu.edu.cn; ya.cheng@siom.ac.cn

[1]The Extreme Optoelectromechanics Laboratory (XXL), School of Physics and Electronic Science, East China Normal University, Shanghai 200241, China
[2]State Key Laboratory of Precision Spectroscopy, East China Normal University, Shanghai 200062, China
[3]State Key Laboratory of High Field Laser Physics and CAS Center for Excellence in Ultra-intense Laser Science, Shanghai Institute of Optics and Fine Mechanics (SIOM), Chinese Academy of Sciences (CAS), Shanghai 201800, China
[4]Collaborative Innovation Center of Extreme Optics, Shanxi University, Taiyuan 030006, China.
[5]Collaborative Innovation Center of Light Manipulations and Applications, Shandong Normal University, Jinan 250358, People's Republic of China
[6]Hefei National Laboratory, Hefei 230088, China
[7]Joint Research Center of Light Manipulation Science and Photonic Integrated Chip of East China Normal University and Shandong Normal University, East China Normal University, Shanghai 200241, China

**Keywords**: thin film lithium niobate, coherent beam combination, photonic integrated circuits, waveguide amplifiers

**Abstract:** We demonstrate on-chip coherent beam combination of two waveguide amplifiers on $Er^{3+}$-doped thin film lithium niobate (Er: TFLN) platform. Our device is built based on an electro-optic modulator fabricated on Er: TFLN. The output power of the coherently combined amplifiers is measured as high as 12.9 mW, surpassing that of previous single waveguide amplifiers based on $Er^{3+}$-doped thin film lithium niobate platform.

## 1．Introduction

High power lasers are of vital importance in various fields of research and application. A promising approach for power scaling of lasers and amplifiers is coherent beam combination (CBC) [1-5]. The power available from a single-channel optical path is limited either by thermal damage or nonlinear effects. In the CBC technique, a seed laser source is first split into several beam paths and each beam is amplified to the highest power provided by a single amplifier within an amplifier array. Then, the output beams from the amplifier array are coherently combined into a single beam. The key requirement for CBC is to phase lock optical amplifier array and thus to guarantee maximum power transfer efficiency during the combination [6-10]. Phase modulator with sufficient bandwidth to dynamically compensate the phase excursion of each beam path, such as electro-optic modulator, liquid crystal modulator, or piezo-electric actuator, is of vital importance for stable phase-locking. Meanwhile, in addition to spatial properties, CBC also preserves the spectral properties of the seed laser. In comparison to coherently-combined fiber laser source with well-established passive or active phase-locking methods [11,12], CBC of on-chip laser and amplifier is rarely investigated due to the limited tuning method and response time in integrated photonic platforms. Featuring with the large linear electro-optical coefficients, low optical losses, and broad transparency band, thin film lithium niobate (TFLN) is an ideal platform for precise and ultrafast optical phase modulation [13-17]. In addition, TFLN can also be doped with rare earth ions (REI) to generate optical gain for on-chip waveguide amplifiers [18-26]. Therefore, phase locking and optical amplifying can be realized simultaneously on REI doped TFLN platform.

So far, the TFLN waveguide amplifiers operating at 1 μm and 1.5 μm have been achieved on $Yb^{3+}$-doped TFLN and $Er^{3+}$-doped TFLN, respectively [18-26]. However, the output power of the single waveguide amplifier is limited around 1 milliwatt (mW) due to gain saturation and absorption losses. To break such gain saturation and further increase the output power, a straightforward way is to enlarge the mode area of the waveguide amplifier, though multimode guiding could spoil the power conversion efficiency and the beam quality of the output signal. Similarly, increasing the waveguide length is less efficient as well due to the excited state absorption and nonlinear distortion at high pump powers necessitated by the suitable inversion of the full waveguide length. In this regard, coherent beam combining of single-mode waveguide amplifiers is highly preferred for continuous power scaling, due to the suppression of the multimode crosstalks and excess nonlinear excitations.

Here, we demonstrate, for the first time to the best of our knowledge, an on-chip CBC of waveguide amplifiers on the $Er^{3+}$-doped TFLN platform. The CBC chip with the footprint of 3.5 mm × 9 mm is fabricated using the photolithography assisted chemo-mechanical etching (PLACE) technique, including two waveguide amplifiers sandwiched by two 2×2 multimode interferometers (MMI), and an integrated TFLN electro-optic Mach–Zehnder interferometer (MZI) modulator used to adjust the relative phase between the two waveguide amplifiers. In the experiment, we measure a $V_\pi$ of 8.3 V between the two waveguide amplifiers and an optical power extinction ratio (ER) of about 13 dB between modulated maximum and minimum output powers. We achieve an output power of 12.9 mW from the on-chip CBC of waveguide amplifiers, greatly surpassing the powers of previous single waveguide amplifiers in the same $Er^{3+}$-doped TFLN platform [18-23]. Our approach opens the venue to ultra-high-power multi-channel CBC photonic circuits with highly accurate and robust phase locking that enabling a wide range of applications including optical communications, laser cooling, and Lidar.

## 2. Device Design and Fabrication

The on-chip CBC of waveguide amplifiers is fabricated on 500-nm thick X-cut $Er^{3+}$-doped TFLN platform. Figure 1(a) illustrates the schematic diagram of the on-chip CBC of waveguide amplifiers, including two waveguide amplifiers folded in spiral shapes, two 2×2 multimode interferometer (MMI), and ground-signal-ground (GSG) microelectrodes. As shown in the energy level diagram of erbium ion in the bottom right corner of Fig. 1(a), the in-band pumping by the 1480-nm laser is employed due to the reduced quantum defect and the single-mode guiding when pumped at 1480 nm instead of 980 nm. The two 2×2 MMIs are designed to function as wavelength division multiplexed beam splitter (combiner) as shown in Figs. 1(b) and (c), featuring uniform power splitting of both the signal (1530 nm) and pump (1480 nm) lights in a 2×2 MMI. It is noted that the signal (1530 nm) and pump (1480 nm) lights incident from different ports will pass through the cascaded MMI and are emitted from the reverse ports. As shown in Figs. 1(d) and (e), in the experiment, we fabricated a short cascaded MMI structure to achieve a phase shift between the two waveguides connected to the MMI that is as close to 0 as possible. We tested them to obtain good power splitting of both the signal (1530 nm) and pump (1480 nm) lights incident at different ports. Finally, we fabricated our devices on a 500-nm-thick X-cut $Er^{3+}$-doped TFLN on a $SiO_2$/Silicon (4.7 μm/500 μm) substrate (NANOLN). The details about the device fabrication can be found in our previous works [27, 28]. We fabricate densely-packed on-chip CBC waveguide amplifiers of 2.5 cm in length (Fig. 1f) with a compact footprint of only 13.5 mm × 9 mm. Fig. 1(g) indicates two waveguide amplifiers integrated with GSG microelectrodes for phase-locking, the gap between microelectrodes is about 7.6 μm, the size of each microelectrode is 3 mm × 100 μm. Fig. 1(h) shows the enlarged part of the input waveguide of 2×2 MMI, the MMI has a compact footprint of only 492 μm × 13 μm.

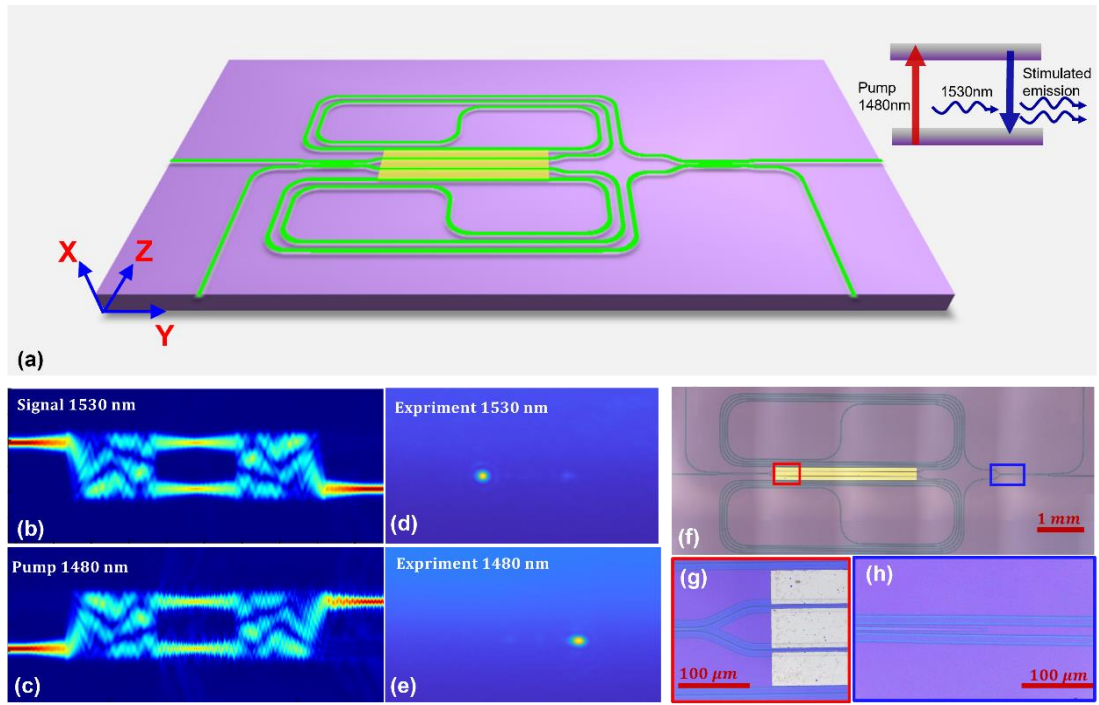

Figure 1. (a) Schematic of the on-chip CBC of $Er^{3+}$ doped waveguide amplifiers pumped by 1480-nm laser diode. Simulated light propagation in the cascaded MMI structure at the wavelength of (b) 1530 nm and (c)1480 nm. Infrared CCD captured mode profile image at the wavelength of (d) 1530 nm and (e)1480 nm of the output of the cascaded MMI. (f) Optical image of on-chip CBC waveguide amplifiers. (g) Optical micrographs of the GSG-microelectrodes for phase tuning and (h) the input waveguides of MMI indicated by the color boxes in (f).

## 3. Characterization and Discussion

We characterized the on-chip CBC of waveguide amplifiers using an optical setup shown in Fig. 2(a). A C-band continuously tunable laser (CTL 1550, TOPTICA Photonics Inc.) is coupled to the port-1 of on-chip CBC waveguide amplifiers using a lensed fiber. In addition, two 1480-nm laser diode (PL-FP-1480-A-A81-SA, LD-PD INC) are coupled to the port-2 and port-3 of on-chip CBC waveguide amplifiers to provide pumping. The polarization state of both the signal and pump lasers are adjusted using in-line fiber polarization controllers (FPC561, Thorlabs Inc.). The output power from port-4 is collected by a lensed fiber with the optical spectrum being monitored by the optical spectrum analyzer (OSA: AQ6375B, YOKOGAWA Inc.). The coupling losses between the lensed fiber and the TFLN waveguides is about 10 dB for both the pump and signal wavelengths. The configuration of the integrated electrodes in the electro-optic modulator is shown in the inset of Fig. 2(a). A user-defined voltage signal provided by the arbitrary function generator (AFG) was applied to the GSG electrodes through the radio frequency (RF) probe. The optical image of an intensely pumped on-chip CBC waveguide amplifiers chip which is butt-coupled with four lensed fibers and controlled by a RF probe is shown in Fig. 2(b). The simulated optical TE mode profile and electric field distribution formed by the GSG electrodes is also shown in Fig. 2(c), the tightly confined optical mode enables us to position metal RF electrodes in proximity to the LN waveguide, resulting in the facilitation of low operational voltages.

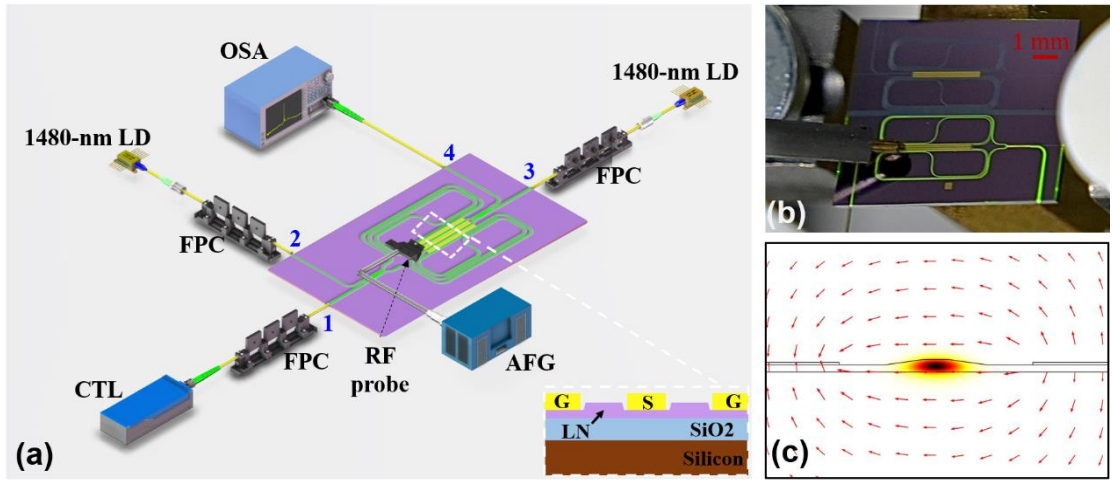

Figure 2. (a) Schematic of the setup for characterizing the on-chip CBC of waveguide amplifiers, Inset shows a schematic of the device cross-section of electro-optic modulator. (CTL: continuously tunable laser; FPC: fiber polarization controller; LD: laser diode; AFG: arbitrary function generator; RF: radio frequency; OSA: optical spectrum analyzer; GSG: ground-signal-ground). (b) optical image of an intensely pumped on-chip CBC of waveguide amplifiers chip butt-coupled with four lensed fibers and controlled by a RF probe. (c) Cross-section view of the simulated optical TE mode profile and RF electrical field (shown by arrows).

Fig. 3(a) shows the output spectra from the on-chip CBC waveguide amplifiers with and without pump laser (red and blue curves). In this case, the sum-power for double-side pumping power injected into the chip is 15 mW, and the signal power is 6 mW. A remarkable enhancement of the output signal power is observed with 1480-nm pump laser. Due to inevitable phase shifts between the two waveguide amplifiers, the 2×2 MMI will deviate from perfect beam combiner with the amplified signal and residual pump mixed in port-3 and port-4. Therefore, it is necessary to apply electrical modulation to compensate for this phase shift and achieve perfect beam combining. A triangular waveform signal with a frequency of 1 Hz and a voltage amplitude of 30 $V_{pp}$ is applied to the 3 mm-long microelectrodes between the two waveguide amplifiers. The power response from port-4 of the CBC chip is shown in Fig. 3(b), and by fitting with the sinusoidal curve, the half-wave voltage is obtained as 8.3 V, corresponding to a $V_\pi \cdot L$ of 2.49 V·cm. It is also observed in Fig. 3(b) that the initial phase of the signal light is perturbed by $\Delta\varphi$ before and after applying the pump laser, most likely due to the photorefractive effect induced by the pump laser [29-31]. Fig. 3(c) indicates that the

ER of signal laser without the pump laser is ~15 dB, and the signal light measured after applying the pump laser results in a power ratio of ~13 dB between the maximum and minimum output powers as shown in Fig. 3(d). The difference in the ER before and after applying the pump laser is also attributed to the photorefractive effect induced by the pump laser. It should be mentioned that the minimum output power at port-4 indicates that the coherently combined signal is preferentially emitted from port-3.

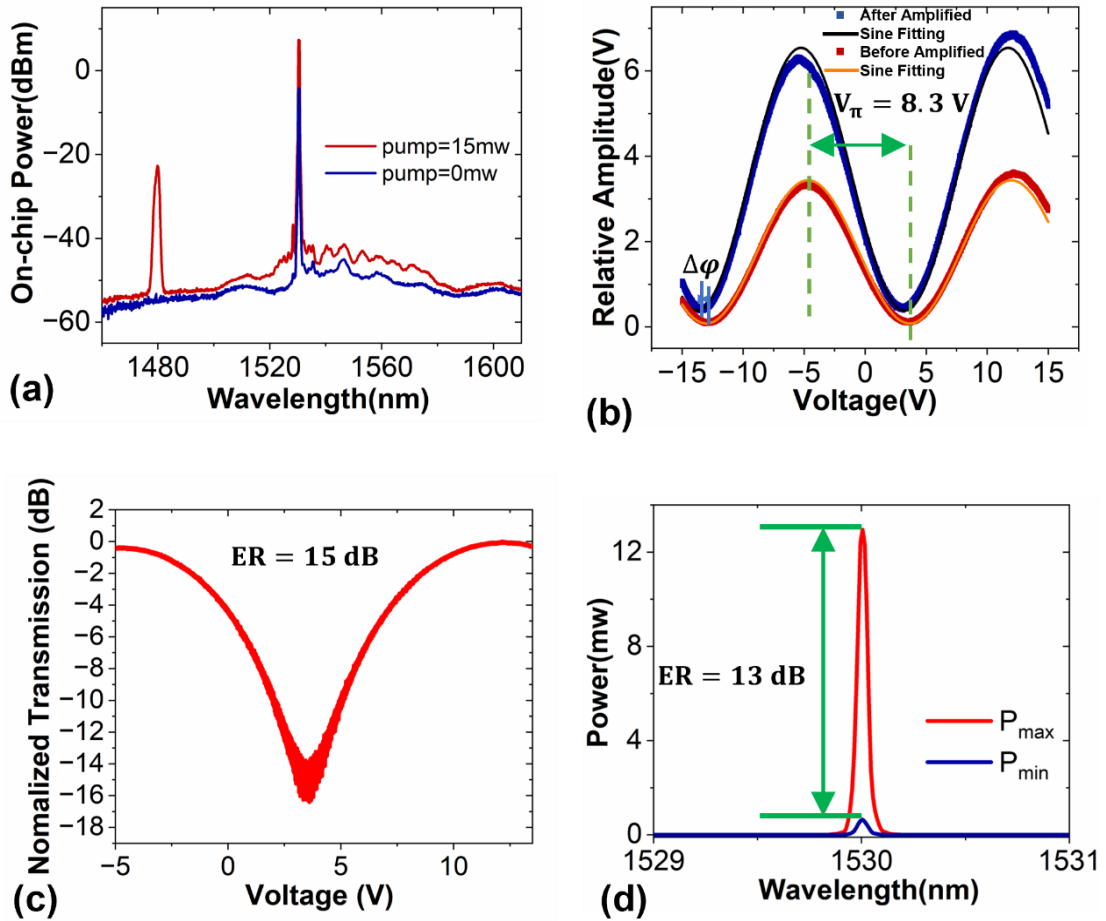

Figure 3. (a) The output spectra from port-4 of the on-chip CBC waveguide amplifiers with and without pump laser (red and blue curves). (b) Electrical modulation of the output signal power by the triangular voltage wave exerted on the 3 mm-long microelectrodes. (c) The measured extinction ratio (ER) is 15 dB without the pump laser. (d) The measured ER of signal light after applying the pump laser is 13 dB.

To further investigate the output performance of the CBC waveguide amplifiers, the amplified signal powers and corresponding signal enhancement factors are measured at two different signal wavelengths and injected signal powers. The signal enhancement factors measured at 1530 nm and 1550 nm with 0.1 mW input signal powers are shown in Fig. 4(a), showing a higher gain of the 1530 nm signal compared to that of the 1550 nm signal. Since the 1530 nm signal is close to the absorption and emission peaks (around 1531.6 nm) of erbium ions, the observed higher gain implies a uniform and high inversion of erbium ions pumped by the 1480 nm laser along the full waveguide length without significant reabsorption losses. The on-chip output powers of the CBC waveguide amplifiers as a function of the pump power are further plotted in Fig. 4(b) for two different input signal powers. The maximum output signal power reaches 12.9 mW for an input power of 6 mW when pumped at 15.8 mW of the 1480-nm laser, indicating an on-chip net gain of 3 dB and a power conversion efficiency of 38%. The calibrated signal spectrum at the maximum output power is shown in the inset of Fig. 4(b).

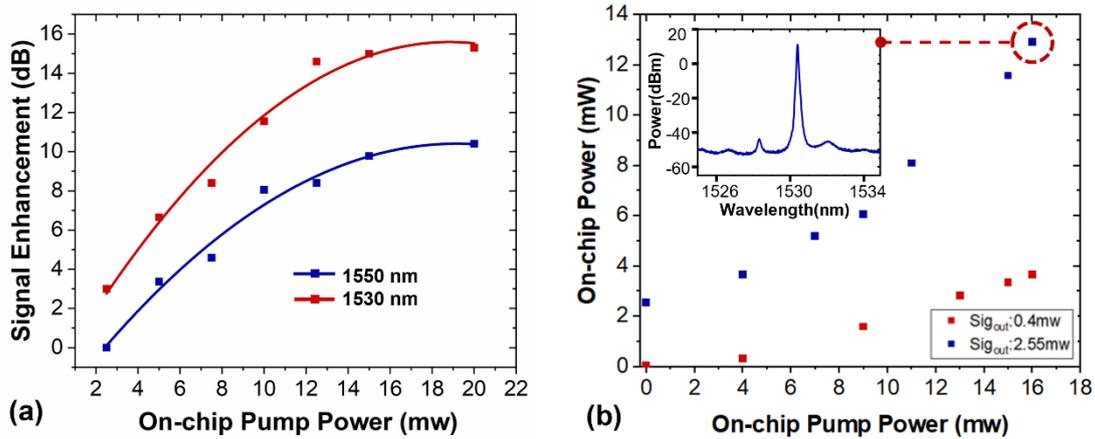

Figure 4. (a) Measured signal enhancement at the signal wavelengths of 1530 nm and 1550 nm on CBC waveguide amplifiers upon pumping at 1480 nm. (b) Measured output signal powers at 1530 nm. The inset shows the calibrated signal spectrum at the maximum output power.

Coherent beam combining is usually applied when the output power from single optical channel is saturated close to thermal damage or significant nonlinear effects. In such cases, the thermal load and nonlinear distortion are already remarkable which will induce a large amount of phase shifts in the output beam and make the coherent beam combining a very challenging task necessitating sophisticated locking protocols. The demonstrated on-chip CBC of $Er^{3+}$-doped waveguide amplifiers are working in a relatively low power regime where a direct phase tuning in the open-loop configuration is sufficient to suppress the phase shift and achieve coherent beam combining. To investigate the internal states of the waveguide amplifiers, an established amplifier model including the steady-state response of erbium ions and the modal distribution of the involved light waves is adopted to simulate the gain performance of the 2.5cm-long waveguide amplifier [32].

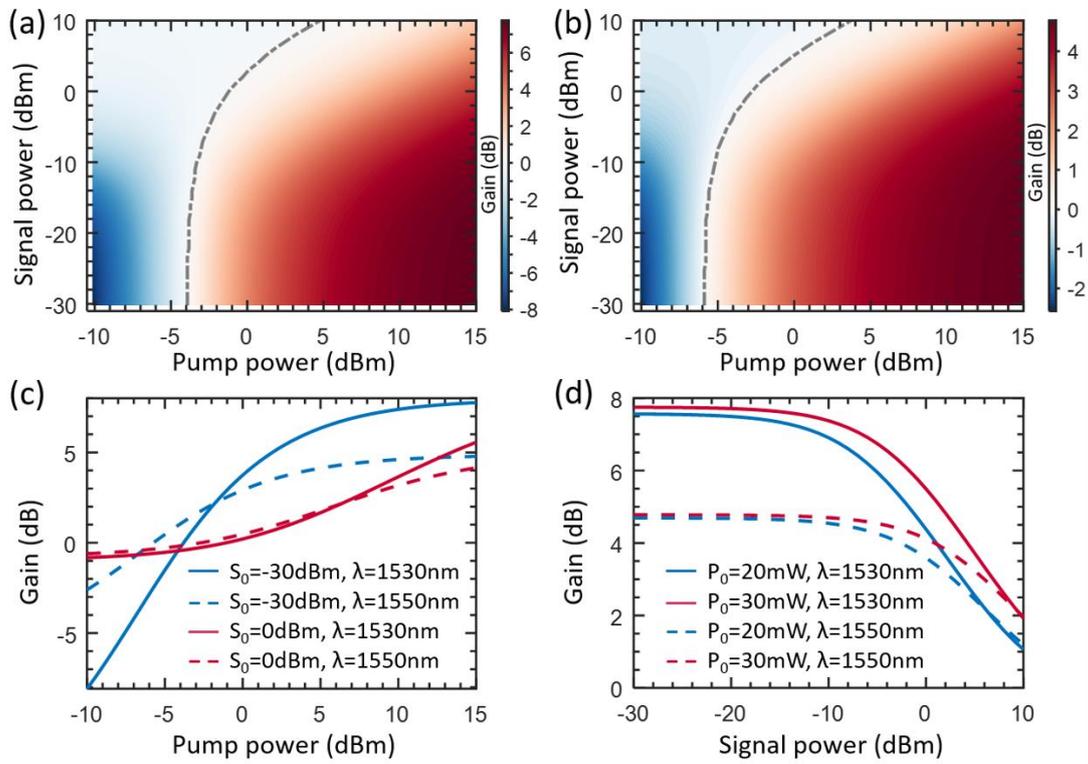

Figure 5. (a-b) The calculated gain values for the signal wavelengths of (a) 1530 nm and (b) 1550 nm. (c) The gain responses for fixed signal powers and increasing pump powers. (d) The gain responses for fixed pump powers and increasing signal powers.

The simulation results are shown in Fig. 5. Gain values calculated for the signal wavelengths of 1530 nm and 1550 nm as the function of input pump and signal powers are depicted in pseudo-colors in Figs. 5(a) and 5(b), respectively. The dashed lines in Figs. 5(a) and 5(b) denote the transparency threshold (0-dB gain) for different input powers. Higher amplifier gain is achieved for high pump powers in the small-signal regime where the population inversion is less affected by the signal amplification. The gain responses at increasing pump (signal) powers for fixed signal (pump) powers are further shown in Figs. 5(c) and 5(d). It can be clearly seen from Fig. 5 that the small-signal gain is higher at 1530nm than 1550nm for high pump powers, while at the high signal power of 1 mW the gain difference between the two wavelengths is greatly reduced due to gain saturation. Meanwhile, it can be noticed from Fig. 5(d) that at both signal wavelengths, the small-signal gain is saturated for high pump powers, while the saturated output powers critically depend on the pump powers. Further simulation results demonstrate deep saturation of amplifier power by increasing waveguide length and pump power, though the coherent beam combining in the high-power regime will greatly suffer from thermal load and nonlinear effects which needs fast phase compensation in a closed-loop configuration.

## 4. Conclusion

We have successfully demonstrated an on-chip coherent beam combining of waveguide amplifiers on X-cut $Er^{3+}$-doped TFLN platform. The CBC waveguide amplifier chip is fabricated by PLACE technique, supporting a coherently combined output power of 12.9 mW which is the highest power among the reported waveguide amplifiers based on $Er^{3+}$-doped TFLN platform. Experimental characterizations reveal the reliable phase shift compensation between two waveguide amplifiers by on-chip TFLN electro-optical modulators with integrated microelectrodes. This work opens the way to unprecedented power scaling of on-chip light sources by robust beam combining through compact and tunable photonic integrated circuits, promising a broad spectrum of applications in optical communication, laser manipulation, light based remote sensing and ranging.

**Acknowledgements**
National Key R&D Program of China (2019YFA0705000, 2022YFA1404600, 2022YFA1205100), National Natural Science Foundation of China (Grant Nos. 12274133, 12004116, 12104159, 12192251, 11933005, 12134001, 61991444, 12204176), Science and Technology Commission of Shanghai Municipality